\documentclass[pra,twocolumn,showpacs]{revtex4-1}
\usepackage{amsmath,amssymb,bm,graphicx,hyperref}

\renewcommand\d{\partial}
\newcommand\up{\uparrow}
\newcommand\down{\downarrow}
\newcommand\eps{\epsilon}
\newcommand\0{\bm{0}}
\renewcommand\k{{\bm{k}}}
\renewcommand\r{{\bm{r}}}
\renewcommand\Re{\mathrm{Re}}
\newcommand\T{\mathrm{T}}
\newcommand\Tr{\mathrm{Tr}}
\newcommand\vF{v_\mathrm{F}}
\newcommand\eff{\mathrm{eff}}
\newcommand\sgn{\mathrm{sgn}}
\newcommand\G{\mathcal{G}}

\begin{document}

\title{Vacuum polarization of graphene with a supercritical Coulomb impurity:\\
Low-energy universality and discrete scale invariance}

\author{Yusuke Nishida}
\affiliation{Department of Physics, Tokyo Institute of Technology,
Ookayama, Meguro, Tokyo 152-8551, Japan}

\date{May 2014}

\begin{abstract}
 We study massless Dirac fermions in a supercritical Coulomb potential
 with the emphasis on that its low-energy physics is universal and
 parametrized by a single quantity per supercritical angular momentum
 channel.  This low-energy parameter with the dimension of length is
 defined only up to multiplicative factors and thus each supercritical
 channel exhibits the discrete scale invariance.  In particular, we show
 that the induced vacuum polarization has a power-law tail whose
 coefficient is a sum of log-periodic functions with respect to the
 distance from the potential center.  This coefficient can also be
 expressed in terms of the energy and width of so-called atomic collapse
 resonances.  Our universal predictions on the vacuum polarization and
 its relationship to atomic collapse resonances shed new light on the
 longstanding fundamental problem of quantum electrodynamics and can in
 principle be tested by graphene experiments with charged impurities.
\end{abstract}


\maketitle

\section{Introduction}
Fate of vacuum in a strong Coulomb potential produced by a heavy atomic
nucleus is a fundamental problem of quantum electrodynamics and has been
the subject of a long study~\cite{Zeldovich:1971}.  One of the physical
quantities that have attracted particular interest is the induced vacuum
polarization~\cite{Wichmann:1954,Brown:1974,Gyulassy:1974,Milshtein:1982,Greiner:1985}.
It is well known that there are two distinct regimes, subcritical
$Z<\alpha^{-1}$ and supercritical $Z>\alpha^{-1}$, depending on the
nuclear charge $Z$ relative to the reciprocal of the fine structure
constant
$\alpha=e^2/(4\pi\varepsilon_0\hbar c)\approx1/137$~\cite{Pomeranchuk:1945}.
Although the resulting physics can be qualitatively different in the two
regimes~\cite{Greiner:1985}, intriguing phenomena caused by the
supercritical Coulomb potential remain elusive because of the absence of
such superheavy atomic nuclei.

The situation has changed since 2004 when a graphene was successfully
isolated~\cite{Novoselov:2004,Novoselov:2005}, which realizes massless
Dirac fermions in two dimensions with the Fermi velocity
$\vF\approx10^6$\,m/s~\cite{Castro-Neto:2009}.  Because the
corresponding ``fine structure constant'' is as large as
$e^2/(4\pi\varepsilon_0\hbar\vF)\approx2$, a cluster of charged
impurities placed on graphene can produce the supercritical Coulomb
potential and thus the resulting intriguing phenomena are now within
experimental reach~\cite{Wang:2012,Wang:2013,Luican-Mayer:2014}.
Theoretically, the vacuum polarization of graphene has been studied
intensively in both subcritical and supercritical
regimes~\cite{Katsnelson:2006,Pereira:2007,Shytov:2007a,Biswas:2007,Fogler:2007,Terekhov:2008,Kotov:2008a,Kotov:2008b,Pereira:2008,Kotov:2012}.
However, in our opinion, the vacuum polarization induced by the
supercritical Coulomb potential has not been understood completely even
without electron-electron interaction.  The objective of this paper is
to shed new light on this longstanding fundamental problem.

Noninteracting massless Dirac fermions in two dimensions subject to a
Coulomb potential are described by the Dirac equation:
\begin{align}\label{eq:dirac}
 \left(-i\bm\d\cdot\bm\sigma-\frac{g}{r}\right)
 \Psi_{\eps j}(\r) = \eps\,\Psi_{\eps j}(\r).
\end{align}
Here $\bm\d\cdot\bm\sigma\equiv\d_x\sigma_x+\d_y\sigma_y$ is the kinetic
term, $\eps\equiv E/\hbar\vF$ is the normalized energy, and
$g\equiv Ze^2/(4\pi\varepsilon_0\hbar\vF)$ is the dimensionless coupling
constant with $-e$ and $Ze$ being the electron and impurity charges,
respectively.  Because the Coulomb potential is circularly symmetric,
the wave function $\Psi_{\eps j}(\r)$ can be chosen as an eigenfunction
of the conserved total angular momentum;
$j=\pm1/2,\pm3/2,\dots$~\cite{Pereira:2007,Shytov:2007a}.  Accordingly,
the vacuum polarization electron density is formally expressed as
\begin{align}\label{eq:density}
 n(\r) = \frac12\sum_{\eps<0}\sum_j|\Psi_{\eps j}(\r)|^2
 - \frac12\sum_{\eps>0}\sum_j|\Psi_{\eps j}(\r)|^2
\end{align}
assuming appropriate normalization and
regularization~\cite{Maksym:2013}.  While an explicit calculation will
be performed below, the functional form of $n(\r)$ can be deduced only
from symmetry and dimensional analysis.  In particular, because of the
absence of intrinsic scale in the subcritical regime $|g|<1/2$, the
induced electron density has to be in a scale invariant form and the
only possibility is
\begin{align}
 n(\r) = N_0\delta(\r).
\end{align}
We note that another apparently possible form $\sim1/r^2$ is not
compatible with the scale invariance because its Fourier transform
generates a logarithm which requires some scale~\cite{Biswas:2007}.
Therefore, the vacuum polarization induced by the subcritical Coulomb
potential is localized at the potential
center~\cite{Pereira:2007,Shytov:2007a,Biswas:2007,Terekhov:2008} and
the analytical expression for the induced electron number $N_0$ was
obtained in Ref.~\cite{Terekhov:2008}.

On the other hand, in the supercritical regime $|g|>1/2$, the
stronger singularity of the Coulomb potential at the origin has to be
regularized, for example, by allowing a finite size for charged
impurity~\cite{Pomeranchuk:1945}.  However, as long as low-energy
physics is concerned, all different regularization can be parametrized
by a single quantity $r_j^*$ per supercritical angular momentum channel
$|j|<|g|$ through a boundary condition on the wave function at the
origin~\cite{Case:1950}.  As we will find in
Eq.~(\ref{eq:supercritical}), this low-energy parameter $r_j^*$ with the
dimension of length is defined only up to multiplicative factors of
$e^{\pi/\sqrt{g^2-j^2}}$.  As a consequence, each supercritical channel
exhibits the discrete scale invariance and the induced electron density
now has a form
\begin{align}\label{eq:tail}
 n(\r) = N_0\delta(\r) + \sum_{|j|<|g|}\frac{F_j(r/r_j^*)}{r^2},
\end{align}
where $F_j(r/r_j^*)$ is an unknown but log-periodic function satisfying
$F_j(r/r_j^*)=F_j(e^{n\pi/\sqrt{g^2-j^2}}r/r_j^*)$ with $n$ being an
arbitrary integer.  Therefore, the vacuum polarization induced by the
supercritical Coulomb potential has a power-law tail
$\sim1/r^2$~\cite{Pereira:2007,Shytov:2007a} whose coefficient is a sum
of log-periodic functions with respect to the distance from the
potential center.  Although the coefficient of the power-law tail was
considered to be a constant in Refs.~\cite{Pereira:2007,Shytov:2007a},
our explicit calculation will show that $F_j(r/r_j^*)$ is the universal
log-periodic function presented in Eq.~(\ref{eq:coefficient}).

\section{Green's function}
In order to determine the coefficient of the power-law tail in the
induced electron density (\ref{eq:tail}), it is more convenient to
employ a Green's function method rather than directly dealing with the
wave function in
Eq.~(\ref{eq:dirac})~\cite{Wichmann:1954,Brown:1974,Gyulassy:1974,Milshtein:1982,Greiner:1985,Terekhov:2008}.
The single-particle Green's function $G(\eps;\r,\r')$ defined for an
arbitrary $\eps\in\mathbb{C}$ is a solution to
\begin{align}\label{eq:inhomo}
 \left(\eps+i\bm\d\cdot\bm\sigma+\frac{g}{r}-m\sigma_z\right)
 G(\eps;\r,\r') = \delta(\r-\r')\openone,
\end{align}
where the mass term $m\sigma_z$ is introduced to serve as an infrared
cutoff and will be set to zero at the end of
calculations~\cite{Brown:1974,Terekhov:2008}.  By substituting its
partial-wave expansion
\begin{align}\label{eq:partial-wave}
 G(\eps;\r,\r') &= \sum_{j=-\infty}^\infty
 \begin{pmatrix}
  e^{i(j-1/2)\theta} & 0 \\
  0 & ie^{i(j+1/2)\theta}
 \end{pmatrix}
 \frac{\G_j(\eps;r,r')}{2\pi\sqrt{rr'}} \notag\\
 &\quad \times
 \begin{pmatrix}
  e^{-i(j-1/2)\theta'} & 0 \\
  0 & -ie^{-i(j+1/2)\theta'}
 \end{pmatrix}
\end{align}
into Eq.~(\ref{eq:inhomo}) as well as the polar coordinate
representation of the $\delta$ function~\cite{MathWorld}, we find the
radial Green's function $\G_j(\eps;r,r')$ to satisfy
\begin{align}\label{eq:inhomo_radial}
 \begin{pmatrix}
  \eps+\frac{g}{r}-m & -\d_r-\frac{j}{r} \\
  \d_r-\frac{j}{r} & \eps+\frac{g}{r}+m
 \end{pmatrix}
 \G_j(\eps;r,r') &= \delta(r-r')\openone,
\end{align}
where $j=\pm1/2,\pm3/2,\dots$ is the total angular momentum quantum
number.

The analytical expression for $\G_j(\eps;r,r')$ can be obtained in a
similar way to the corresponding problem in three
dimensions~\cite{Wichmann:1954,Gyulassy:1974,Greiner:1985}.  We first
set
\begin{align}\label{eq:green_radial}
 \G_j(\eps;r,r') &= \theta(r-r')\psi_{\eps j}^>(r)[\psi_{\eps j}^<(r')]^\T \notag\\
 &\quad + \theta(r'-r)\psi_{\eps j}^<(r)[\psi_{\eps j}^>(r')]^\T
\end{align}
with 
$\psi_{\eps j}^>(r)=[\psi_{\eps j\up}^>(r),\psi_{\eps j\down}^>(r)]^\T$
and
$\psi_{\eps j}^<(r)=[\psi_{\eps j\up}^<(r),\psi_{\eps j\down}^<(r)]^\T$
being solutions to the radial Dirac equation:
\begin{align}\label{eq:dirac_radial}
 \begin{pmatrix}
  \eps+\frac{g}{r}-m & -\d_r-\frac{j}{r} \\
  \d_r-\frac{j}{r} & \eps+\frac{g}{r}+m
 \end{pmatrix}
 \psi_{\eps j}(r) = 0.
\end{align}
These two solutions have to be normalized as
\begin{align}\label{eq:normalization}
 \psi_{\eps j\up}^>(r)\psi_{\eps j\down}^<(r)
 - \psi_{\eps j\down}^>(r)\psi_{\eps j\up}^<(r) = 1
\end{align}
to satisfy Eq.~(\ref{eq:inhomo_radial}) and, in addition, have to be
chosen so that the radial Green's function $\G_j(\eps;r,r')$ satisfies
appropriate boundary conditions at $r\to\infty$ and $r\to0$ with $r'$
fixed.  It is obvious from its expression (\ref{eq:green_radial}) that
the long-distance limit is controlled by $\psi_{\eps j}^>(r)$ which has
to be bounded at $r\to\infty$:
\begin{align}\label{eq:long-distance}
 \lim_{r\to\infty}\left|\psi_{\eps j}^>(r)\right|<\infty.
\end{align}
On the other hand, the short-distance limit is controlled by
$\psi_{\eps j}^<(r)$ whose boundary condition at $r\to0$ requires
different treatment for subcritical and supercritical angular momentum
channels~\cite{Pereira:2007,Shytov:2007a}.

In a subcritical angular momentum channel $|j|>|g|$, the radial Dirac
equation (\ref{eq:dirac_radial}) admits regular and singular solutions
$\psi_{\eps j}^<(r)\to r^{\pm\bar\gamma}(j\pm\bar\gamma,g)^\T$ at
$r\to0$ with the real exponent $\bar\gamma\equiv\sqrt{j^2-g^2}$.
Because low-energy physics is dominated by the regular solution, the
relevant boundary condition is to impose
\begin{subequations}\label{eq:short-distance}
\begin{align}\label{eq:subcritical}
 \lim_{r\to0}\psi_{\eps j}^<(r) \propto r^{\bar\gamma}
 \begin{pmatrix}
  j+\bar\gamma \\ g
 \end{pmatrix}.
\end{align}
On the other hand, in a supercritical angular momentum channel
$|j|<|g|$,  the above two solutions become oscillatory
$\psi_{\eps j}^<(r)\to r^{\pm i\gamma}(j\pm i\gamma,g)^\T$ at $r\to0$ with
the imaginary exponent $i\gamma\equiv i\sqrt{g^2-j^2}$.  Because both
solutions are now equally important to low-energy physics, the general
solution becomes their superposition which is uniquely specified by
imposing the boundary condition:
\begin{align}\label{eq:supercritical}
 \lim_{r\to0}\psi_{\eps j}^<(r)
 &\propto \biggl(\frac{r}{r_j^*}\biggr)^{i\gamma}
 \begin{pmatrix}
  j+i\gamma \\ g
 \end{pmatrix}
 - \biggl(\frac{r}{r_j^*}\biggr)^{-i\gamma}
 \begin{pmatrix}
  j-i\gamma \\ g
 \end{pmatrix}.
\end{align}
\end{subequations}
We thus find that the solution in each supercritical channel is
parametrized by a single quantity $r_j^*>0$ with the dimension of
length.  An important point, which seems not to be fully appreciated in
previous studies, is that the emergent low-energy parameter $r_j^*$ is
defined only up to multiplicative factors of $e^{\pi/\gamma}$, i.e.,
$e^{n\pi/\gamma}r_j^*$ with $n$ being an arbitrary integer corresponds
to the same physics.  As a consequence, each supercritical channel
exhibits the discrete scale invariance, which also emerges in
nonrelativistic one-body~\cite{Case:1950}, two-body~\cite{Nishida:2012},
and three-body~\cite{Efimov:1970} problems and can be viewed as a
manifestation of the quantum scale anomaly and the renormalization group
limit cycle~\cite{Gorsky:2014,Bulycheva:2014}.

It is then straightforward to construct $\psi_{\eps j}^>(r)$ and
$\psi_{\eps j}^<(r)$ satisfying the above required conditions
(\ref{eq:normalization})--(\ref{eq:short-distance}) from the following
two linearly independent solutions to the radial Dirac equation
(\ref{eq:dirac_radial})~\cite{Novikov:2007}:
\begin{align}
 \psi_{\eps j}^{(1,2)}(r) =
 \begin{pmatrix}
  \phantom{\sgn(m)}\sqrt{m+\eps}
  \left[u_{\eps j}^{(1,2)}(r)+v_{\eps j}^{(1,2)}(r)\right] \\
  \sgn(m)\sqrt{m-\eps}\left[u_{\eps j}^{(1,2)}(r)-v_{\eps j}^{(1,2)}(r)\right]
 \end{pmatrix}
\end{align}
with
\begin{subequations}\label{eq:solution1}
\begin{align}
 u_{\eps j}^{(1)}(r) &= \frac{g\frac{\eps}{\kappa}+\bar\gamma}{2\kappa}
 \frac{\Gamma(1-g\frac{\eps}{\kappa}+\bar\gamma)}{\Gamma(1+2\bar\gamma)} \notag\\
 &\quad \times (2\kappa r)^{\bar\gamma}e^{-\kappa r}
 M(-g\tfrac{\eps}{\kappa}+\bar\gamma,1+2\bar\gamma,2\kappa r), \\
 v_{\eps j}^{(1)}(r) &= \frac{j-g\frac{m}{\kappa}}{2\kappa}
 \frac{\Gamma(1-g\frac{\eps}{\kappa}+\bar\gamma)}{\Gamma(1+2\bar\gamma)} \notag\\
 &\quad \times (2\kappa r)^{\bar\gamma}e^{-\kappa r}
 M(1-g\tfrac{\eps}{\kappa}+\bar\gamma,1+2\bar\gamma,2\kappa r)
\end{align}
\end{subequations}
and
\begin{subequations}\label{eq:solution2}
\begin{align}
 u_{\eps j}^{(2)}(r) &= \frac{-1}{j-g\frac{m}{\kappa}}
 (2\kappa r)^{\bar\gamma}e^{-\kappa r}
 U(-g\tfrac{\eps}{\kappa}+\bar\gamma,1+2\bar\gamma,2\kappa r), \\
 v_{\eps j}^{(2)}(r) &= (2\kappa r)^{\bar\gamma}e^{-\kappa r}
 U(1-g\tfrac{\eps}{\kappa}+\bar\gamma,1+2\bar\gamma,2\kappa r).
\end{align}
\end{subequations}
Here $\kappa\equiv\sqrt{m^2-\eps^2}$ is introduced, $M(a,b,z)$ and
$U(a,b,z)$ are the confluent hypergeometric functions of the first and
second kind~\cite{Abramowitz:1965}, respectively, and $\bar\gamma$ for
$|j|>|g|$ should be understood as $\bar\gamma=i\gamma$ when $|j|<|g|$.
By setting
\begin{align}
 \psi_{\eps j}^>(r) = \psi_{\eps j}^{(2)}(r)
\end{align}
and
\begin{align}
 \psi_{\eps j}^<(r) = \psi_{\eps j}^{(1)}(r) + C_{\eps j}\psi_{\eps j}^{(2)}(r),
\end{align}
the normalization condition (\ref{eq:normalization}) and the
long-distance boundary condition (\ref{eq:long-distance}) are
satisfied.  The so far arbitrary prefactor $C_{\eps j}$ is chosen to
satisfy the short-distance boundary condition (\ref{eq:short-distance}),
which leads to $C_{\eps j}=0$ for a subcritical angular momentum channel
$|j|>|g|$ and
\begin{align}\label{eq:prefactor}
 C_{\eps j} = \frac{-\frac{j-g\frac{m}{\kappa}}{2\kappa}
 \frac{j-g\frac{m+\eps}{\kappa}-i\gamma}{(2\kappa r_j^*)^{-i\gamma}}
 \frac{\Gamma(1-g\frac{\eps}{\kappa}+i\gamma)}{\Gamma(2i\gamma)}}
 {\frac{j-g\frac{m+\eps}{\kappa}+i\gamma}{(2\kappa r_j^*)^{i\gamma}}
 \frac{\Gamma(1+2i\gamma)}{\Gamma(1-g\frac{\eps}{\kappa}+i\gamma)}
 - \frac{j-g\frac{m+\eps}{\kappa}-i\gamma}{(2\kappa r_j^*)^{-i\gamma}}
 \frac{\Gamma(1-2i\gamma)}{\Gamma(1-g\frac{\eps}{\kappa}-i\gamma)}}
\end{align}
for a supercritical angular momentum channel $|j|<|g|$.  Accordingly,
the single-particle Green's function (\ref{eq:partial-wave}) is
completely determined.  For later use, we decompose the radial Green's
function into two parts as
\begin{align}\label{eq:decomposition}
 \G_j(\eps;r,r') = \G_j^0(\eps;r,r') + \delta\G_j(\eps;r,r')
\end{align}
with
\begin{align}
 \G_j^0(\eps;r,r') &\equiv \theta(r-r')\psi_{\eps j}^{(2)}(r)
 [\psi_{\eps j}^{(1)}(r')]^\T \notag\\
 &\quad + \theta(r'-r)\psi_{\eps j}^{(1)}(r)[\psi_{\eps j}^{(2)}(r')]^\T
\end{align}
and
\begin{align}
 \delta\G_j(\eps;r,r') \equiv C_{\eps j}\psi_{\eps j}^{(2)}(r)
 [\psi_{\eps j}^{(2)}(r')]^\T,
\end{align}
where the low-energy parameter $r_j^*$ with the dimension of length
appears only in the latter through $C_{\eps j}$.

\section{Vacuum polarization}
We are now ready to evaluate the vacuum polarization electron density
(\ref{eq:density}), which is expressed in terms of the single-particle
Green's function (\ref{eq:partial-wave}) as
\begin{align}\label{eq:bare}
 \tilde n(\r) = \sum_{j=-\infty}^\infty\int_{-i\infty}^{i\infty}\!
 \frac{d\eps}{2\pi i}\frac{\Tr\left[\G_j(\eps;\r,\r)\right]}{2\pi r},
\end{align}
where the contour of the integration over $\eps$ is deformed to coincide
with the imaginary
axis~\cite{Wichmann:1954,Brown:1974,Gyulassy:1974,Milshtein:1982,Greiner:1985,Terekhov:2008}.
This formal expression contains divergence which has to be renormalized.
In order to separate out the divergent part from the convergent part, we
use the decomposition (\ref{eq:decomposition}) to rewrite the bare
electron density (\ref{eq:bare}) as
$\tilde n(\r)=\tilde n_0(\r)+\sum_{|j|<|g|}\delta\tilde n_j(\r)$ with
\begin{align}
 \tilde n_0(\r) \equiv \sum_{j=-\infty}^\infty\int_{-i\infty}^{i\infty}\!
 \frac{d\eps}{2\pi i}\frac{\Tr\left[\G_j^0(\eps;r,r)\right]}{2\pi r}
\end{align}
and
\begin{align}
 \delta\tilde n_j(\r) \equiv \int_{-i\infty}^{i\infty}\!
 \frac{d\eps}{2\pi i}\frac{\Tr\left[\delta\G_j(\eps;r,r)\right]}{2\pi r},
\end{align}
which have to be treated separately.  The first part $\tilde n_0(\r)$ is
divergent and thus needs the renormalization by requiring the total
induced electron number to
vanish~\cite{Wichmann:1954,Brown:1974,Milshtein:1982,Terekhov:2008}.
Technically, this renormalization can be performed by considering the
Fourier transform of the bare electron density
$\tilde\nu_0(\k)=\int\!d\r\,e^{-i\k\cdot\r}\tilde n_0(\r)$ with an
ultraviolet cutoff $|\eps|<\Lambda$ and then introducing the
renormalized quantity by
$\nu_0(\k)=\lim_{\Lambda\to\infty}\left[\tilde\nu_0(\k)-\lim_{k\to0}\tilde\nu_0(\k)\right]$
to satisfy the required neutrality condition $\nu_0(\0)=0$.  Because the
mass $m$ is the only dimensionful parameter existing in
$\G_j^0(\eps;r,r)$, the resulting dimensionless function $\nu_0(\k)$ can
depend only on the ratio $k/m$.  Accordingly, it becomes just a constant
in the massless limit,
$\lim_{m\to0}\nu_0(\k)=N_0$~\cite{Biswas:2007,Terekhov:2008}, whose
inverse Fourier transform gives the cutoff ($m,\Lambda$) independent
renormalized electron density
$n_0(\r)=\int\!d\k/(2\pi)^2e^{i\k\cdot\r}\lim_{m\to0}\nu_0(\k)$ as
\begin{align}\label{eq:renormalized1}
 n_0(\r) = N_0\delta(\r),
\end{align}
which does not contribute to the power-law tail.

\begin{figure}[t]
 \includegraphics[width=0.95\columnwidth,clip]{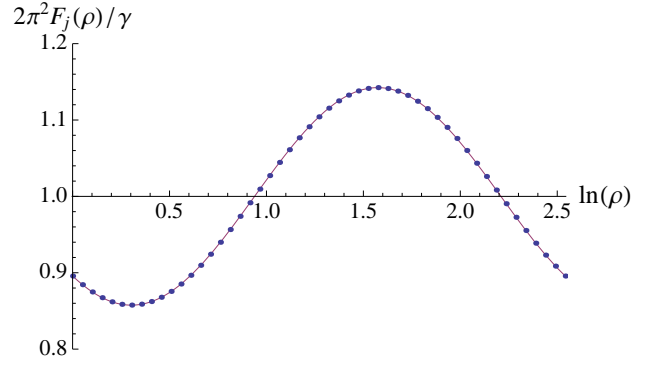}
 \caption{Universal log-periodic function $2\pi^2F_j(\rho)/\gamma$
 obtained in Eq.~(\ref{eq:coefficient}) as a function of $\ln(\rho)$.
 One period in a range $0\leq\ln(\rho)\leq\pi/\gamma\approx2.55$ is
 shown for $j=1/2$ and $g=4/3$.  Dots are the exact result and the solid
 curve is a fit with a single sine function;
 $2\pi^2F_j(\rho)/\gamma\simeq1-0.142\sin[2\gamma\ln(\rho)+0.821]$.
 \label{fig:plot}}
\end{figure}

On the other hand, the second part $\delta\tilde n_j(\r)$ is convergent
and thus does not need the renormalization because its integrand
$\delta\G_j(\eps;r,r)$ decreases exponentially at $\kappa r\to\infty$
[see Eq.~(\ref{eq:solution2})].  Accordingly, its contribution to the
induced electron density
$\delta n_j(\r)=\lim_{m\to0}\delta\tilde n_j(\r)$ can be directly
evaluated by taking the massless limit and is found to have the
power-law form
\begin{align}\label{eq:renormalized2}
 \delta n_j(\r) = \frac{F_j(r/r_j^*)}{r^2}
\end{align}
with the dimensionless coefficient given by
\begin{align}\label{eq:coefficient}
 & F_j(r/r_j^*) = \frac{\gamma}{2\pi^2}\,\Re\int_0^\infty\!dz\,
 \frac{\Gamma(1-ig+i\gamma)\Gamma(1-ig-i\gamma)}
 {\Gamma(1+2i\gamma)\Gamma(1-2i\gamma)} \notag\\
 &\times \left[\frac{1+\frac{(j-ig+i\gamma)\Gamma(1+2i\gamma)\Gamma(1-ig-i\gamma)}
 {(j-ig-i\gamma)\Gamma(1-2i\gamma)\Gamma(1-ig+i\gamma)}
 \bigl(\frac{r}{zr_j^*}\bigr)^{2i\gamma}}
 {1-\frac{(j-ig+i\gamma)\Gamma(1+2i\gamma)\Gamma(1-ig-i\gamma)}
 {(j-ig-i\gamma)\Gamma(1-2i\gamma)\Gamma(1-ig+i\gamma)}
 \bigl(\frac{r}{zr_j^*}\bigr)^{2i\gamma}}\right] \notag\\
 &\times e^{-z}U(-ig+i\gamma,1+2i\gamma,z)U(1-ig-i\gamma,1-2i\gamma,z).
\end{align}
By summing up all the contributions from Eqs.~(\ref{eq:renormalized1})
and (\ref{eq:renormalized2}), we obtain the renormalized electron
density as
\begin{align}
 n(\r) = n_0(\r) + \sum_{|j|<|g|}\delta n_j(\r),
\end{align}
which establishes the form of the induced electron density presented in
Eq.~(\ref{eq:tail}).  In particular, we find that the coefficient of the
power-law tail (\ref{eq:coefficient}) is log-periodic
$F_j(r/r_j^*)=F_j(e^{n\pi/\gamma}r/r_j^*)$ as it must be because
$e^{n\pi/\gamma}r_j^*$ with $n$ being an arbitrary integer corresponds
to the same physics [see Eq.~(\ref{eq:supercritical})] and also
universal in the sense that all microscopic details are parametrized by
the single quantity $r_j^*>0$ per supercritical angular momentum channel
$|j|<|g|$.  Figure~\ref{fig:plot} shows the obtained universal
log-periodic function in the normalized form $2\pi^2F_j(r/r_j^*)/\gamma$
by taking $j=1/2$ and $g=4/3$ as an example.  Its mean value can be
extracted by replacing the whole expression in the square bracket of
Eq.~(\ref{eq:coefficient}) with $\sgn(g)$, which leads to
\begin{align}\label{eq:average}
 \overline{F_j(r/r_j^*)} = \frac{\gamma}{2\pi^2}\,\sgn(g)
\end{align}
in agreement with the constant coefficient of the power-law tail
considered in Ref.~\cite{Shytov:2007a}.  While $F_j(r/r_j^*)$ and its
mean value (\ref{eq:average}) coincide in the limit $|g|\to\infty$, they
significantly deviate especially when $|g|\simeq|j|$.  We also note that
$F_j(r/r_j^*)$ is odd with respect to $g\to-g$ and $F_{-j}(r/r_{-j}^*)$
is essentially the same function as $F_j(r/r_j^*)$ because their
apparent difference can be absorbed by redefining $r_{-j}^*$.

The phase of the log-periodic oscillation in Eq.~(\ref{eq:coefficient})
is fixed by the nonuniversal parameter $r_j^*$ which depends on
microscopic physics and thus cannot be determined from our perspective
of low-energy effective theory.  However, it can be related to other
physical quantities such as the energy and width of so-called atomic
collapse resonances~\cite{Wang:2013}.  Although bound states cannot be
formed in the massless limit, it was theoretically shown that an
infinite family of resonances emerges in each supercritical angular
momentum channel $|j|<|g|$~\cite{Pereira:2007,Shytov:2007b,Shytov:2009}.
Their energy and width are determined by poles of the single-particle
Green's function (\ref{eq:partial-wave}) in the second Riemann sheet of
the complex $\eps$ plane, which can arise only as poles of the prefactor
$C_{\eps j}$ obtained in Eq.~(\ref{eq:prefactor}).  By substituting
$\kappa\to-\sqrt{m^2-\eps^2}$ and taking the massless limit $m\to0$, we
find an infinite family of complex poles at
\begin{align}
 \eps_j^*  = \frac{i}{2r_j^*}
 \left[\frac{(j-ig+i\gamma)\Gamma(1+2i\gamma)\Gamma(1-ig-i\gamma)}
 {(j-ig-i\gamma)\Gamma(1-2i\gamma)\Gamma(1-ig+i\gamma)}\right]^{\frac1{2i\gamma}}
\end{align}
with multiplicative factors of $e^{\pi/\gamma}$, which leads to
\begin{align}\label{eq:resonance}
 E_j^{(n)} - \frac{i}2\,\Gamma_j^{(n)} = \hbar\vF e^{-n\pi/\gamma}\eps_j^*
\end{align}
as the energy and width of the $n$th atomic collapse resonance.
Therefore, if $r_j^*$ for a given system is determined through the
energy or width of an atomic collapse resonance (\ref{eq:resonance}), we
then have an unambiguous prediction for the power-law tail of the vacuum
polarization electron density (\ref{eq:coefficient}), and vice versa.
Furthermore, the complex expression in the square bracket of
Eq.~(\ref{eq:coefficient}) can be greatly simplified by using the
complex energy in Eq.~(\ref{eq:resonance}) as
\begin{align}\label{eq:relation}
 \left[\,\cdots\,\right] = \left[\frac{1+\bigl(\frac{2\eps_j^*r}{iz}\bigr)^{2i\gamma}}
 {1-\bigl(\frac{2\eps_j^*r}{iz}\bigr)^{2i\gamma}}\right],
\end{align}
which model-independently relates the two intriguing phenomena caused by
the supercritical Coulomb potential, i.e., the vacuum polarization and
the atomic collapse resonances.

\begin{figure}[t]
 \includegraphics[width=0.95\columnwidth,clip]{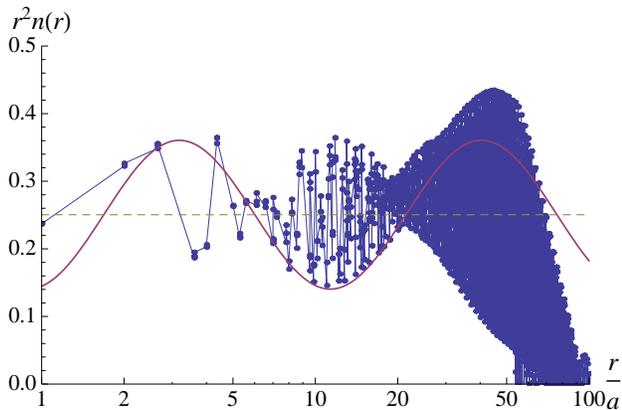}
 \caption{Induced electron density $n(\r)$ for the tight-binding
 Hamiltonian on a honeycomb lattice as a function of $r$ in units of the
 lattice parameter $a$.  Connected dots are numerical data for $g=4/3$
 obtained in Ref.~\cite{Pereira:2007} by the exact diagonalization with
 $124^2$ lattice sites.  The solid curve is a fit based on our
 prediction (\ref{eq:fit}) with the mean value
 $2\gamma/\pi^2\approx0.25$ indicated by the horizontal dashed line.
 \label{fig:comp}}
\end{figure}

Finally, it is worthwhile to compare our prediction with the induced
electron density computed for the tight-binding Hamiltonian on a
honeycomb lattice whose low-energy physics is described by two valley
species of massless Dirac fermions~\cite{Semenoff:1984,DiVincenzo:1984}.
Figure~\ref{fig:comp} shows numerical data for $g=4/3$ obtained in
Ref.~\cite{Pereira:2007} by the exact diagonalization with $124^2$
lattice sites in units of the lattice parameter $a$.  In addition to
rapid oscillations presumably caused by the lattice cutoff, there seems
to be a slow oscillation which should be contrasted with the predicted
log-periodic oscillation.  Because only $j=\pm1/2$ channels are
supercritical for $g=4/3$, our prediction (\ref{eq:tail}) reduces to
$r^2n(\r)=2\sum_{j=\pm1/2}F_j(r/r_j^*)$ including the factor two due to
the valley degeneracy.  Here the universal log-periodic function
$F_j(r/r_j^*)$, as seen in Fig.~\ref{fig:plot}, can be excellently
approximated by a single sine function with a relative error less than
$0.01\%$.  Accordingly, our prediction for the power-law tail of the
induced electron density is expressed as
\begin{align}\label{eq:fit}
 r^2n(\r) \simeq \frac{2\gamma}{\pi^2}
 - A\sin\!\left[2\gamma\ln\!\left(\frac{r}{a}\right)+\varphi\right],
\end{align}
where the amplitude $A$ and the phase $\varphi$ are related to the
unknown two low-energy parameters $r_{\pm1/2}^*/a$.  We find in
Fig.~\ref{fig:comp} that the lower envelope of numerical data in the
intermediate region $a\ll r\ll50\,a$ between the lattice cutoff and the
system radius is well fitted by choosing $A\approx0.11$ and
$\varphi\approx1.85$, which supports the validity of our prediction.

\section{Concluding remarks}
In this paper, we studied massless Dirac fermions in a supercritical
Coulomb potential with the emphasis on that its low-energy physics is
universal and each supercritical angular momentum channel exhibits the
discrete scale invariance.  In particular, we showed that the induced
vacuum polarization has a power-law tail whose coefficient is a sum of
log-periodic functions with respect to the distance from the potential
center.  This coefficient can also be expressed in terms of the energy
and width of so-called atomic collapse resonances.  While these
universal features are explicitly demonstrated only in two dimensions
[see Eqs.~(\ref{eq:renormalized2})--(\ref{eq:relation})], it is
straightforward to extend our present analysis to three dimensions as
well.  Therefore, qualitatively the same features are indeed expected in
the vacuum polarization caused by a superheavy atomic nucleus with
$Z>\alpha^{-1}$ as long as ultraviolet and infrared cutoffs (i.e.,
nuclear charge radius and electron Compton wavelength, respectively) are
well separated compared to log-periodic oscillations, which shed new
light on the longstanding fundamental problem of quantum
electrodynamics.  Furthermore, because low-energy physics of graphene is
described by massless Dirac fermions in two dimensions, our universal
predictions on the vacuum polarization and its relationship to observed
atomic collapse resonances~\cite{Wang:2013} can in principle be tested
experimentally by measuring the induced electron density with scanning
probe microscopy techniques~\cite{Deshpande:2012}.

While the electron-electron interaction has been neglected in our
present analysis, our novel finding on the vacuum polarization may be
useful to develop further insight into the screening of the
supercritical Coulomb impurity in the presence of the electron-electron
interaction.  One possible approach is to write down the self-consistent
renormalization group equation in the same spirit as
Ref.~\cite{Shytov:2007a}:
\begin{align}\label{eq:screening}
 \frac{d_{}Z_\eff(R)}{d\ln R} = -2\pi
 \left[\sum_{|j|<|g|}F_j(R/r_j^*)\right]_{Z\to Z_\eff(R)},
\end{align}
where $Z_\eff(R)\equiv Z-\int_{|\r|<R}d\r\,n(\r)$ multiplied by $e$ is
the total charge within the radius $R$.  Because the right hand side of
Eq.~(\ref{eq:screening}) is negative, the total charge decreases as the
radius increases until the right hand side vanishes, i.e.,
$|g_\eff(R)|\to1/2$, which leads to the screening of the supercritical
charge down to the critical value~\cite{Shytov:2007a}.  Detailed
analysis of the solution to our self-consistent renormalization group
equation (\ref{eq:screening}) shall be deferred to a future work.

\acknowledgments
The author thanks Vitor M.~Pereira for providing numerical data in
Ref.~\cite{Pereira:2007} and related discussions.  This work was
supported by JSPS KAKENHI Grant Number 25887020.


\begin{thebibliography}{99}

\bibitem{Zeldovich:1971}
  Ya.~B.~Zeldovich and V.~S.~Popov,
  Usp.\ Fiz.\ Nauk {\bf 105}, 403 (1971)
  [Sov.\ Phys.\ Usp.\ {\bf 14}, 673 (1972)].

\bibitem{Wichmann:1954}
  E.~Wichmann and N.~M.~Kroll,
  Phys.\ Rev.\ {\bf 96}, 232 (1954);
%
  Phys.\ Rev.\ {\bf 101}, 843 (1956).

\bibitem{Brown:1974}
  L.~S.~Brown, R.~N.~Cahn, and L.~D.~McLerran,
  Phys.\ Rev.\ Lett.\ {\bf 32}, 562 (1974);
%
  Phys.\ Rev.\ Lett.\ {\bf 33}, 1591 (1974);
%
  Phys.\ Rev.\ D {\bf 12}, 581 (1975);
%
  Phys.\ Rev.\ D {\bf 12}, 596 (1975);
%
  Phys.\ Rev.\ D {\bf 12}, 609 (1975).

\bibitem{Gyulassy:1974}
  M.~Gyulassy,
  Phys.\ Rev.\ Lett.\ {\bf 32}, 1393 (1974);
%
  Phys.\ Rev.\ Lett.\ {\bf 33}, 921 (1974);
%
  Nucl.\ Phys.\ A {\bf 244}, 497 (1975).

\bibitem{Milshtein:1982}
  A.~I.~Mil'shtein and V.~M.~Strakhovenko,
  Phys.\ Lett.\ A {\bf 90}, 447 (1982);
%
  Phys.\ Lett.\ A {\bf 92}, 381 (1982);
%
  Zh.\ Eksp.\ Teor.\ Fiz.\ {\bf 84}, 1247 (1983)
  [Sov.\ Phys.\ JETP {\bf 57}, 722 (1983)].

\bibitem{Greiner:1985}
  W.~Greiner, B.~M\"uller, and J.~Rafelski,
  {\it Quantum Electrodynamics of Strong Fields}
  (Springer-Verlag, Berlin, 1985).

\bibitem{Pomeranchuk:1945}
  I.~Ya.~Pomeranchuk and Y.~A.~Smorodinsky,
  J.\ Phys.\ USSR {\bf 9}, 97 (1945).

\bibitem{Novoselov:2004}
  K.~S.~Novoselov, A.~K.~Geim, S.~V.~Morozov, D.~Jiang, Y.~Zhang, S.~V.~Dubonos, I.~V.~Grigorieva, and A.~A.~Firsov,
  Science {\bf 306}, 666 (2004).

\bibitem{Novoselov:2005}
  K.~S.~Novoselov, D.~Jiang, F.~Schedin, T.~J.~Booth, V.~V.~Khotkevich, S.~V.~Morozov, and A.~K.~Geim,
  Proc.\ Natl.\ Acad.\ Sci.\ USA {\bf 102}, 10451 (2005).

\bibitem{Castro-Neto:2009}
  A.~H.~Castro~Neto, F.~Guinea, N.~M.~R.~Peres, K.~S.~Novoselov, and A.~K.~Geim,
  Rev.\ Mod.\ Phys.\ {\bf 81}, 109 (2009).

\bibitem{Wang:2012}
  Y.~Wang, V.~W.~Brar, A.~V.~Shytov, Q.~Wu, W.~Regan, H.-Z.~Tsai, A.~Zettl, L.~S.~Levitov, and M.~F.~Crommie,
  Nat.\ Phys.\ {\bf 8}, 653 (2012).

\bibitem{Wang:2013}
  Y.~Wang, D.~Wong, A.~V.~Shytov, V.~W.~Brar, S.~Choi, Q.~Wu, H.-Z.~Tsai, W.~Regan, A.~Zettl, R.~K.~Kawakami, S.~G.~Louie, L.~S.~Levitov, and M.~F.~Crommie,
  Science {\bf 340}, 734 (2013).

\bibitem{Luican-Mayer:2014}
  A.~Luican-Mayer, M.~Kharitonov, G.~Li, C.-P.~Lu, I.~Skachko, A.-M.~B.~Gon\c{c}alves, K.~Watanabe, T.~Taniguchi, and E.~Y.~Andrei,
  Phys.\ Rev.\ Lett.\ {\bf 112}, 036804 (2014).

\bibitem{Katsnelson:2006}
  M.~I.~Katsnelson,
  Phys.\ Rev.\ B {\bf 74}, 201401(R) (2006).

\bibitem{Pereira:2007}
  V.~M.~Pereira, J.~Nilsson, and A.~H.~Castro~Neto,
  Phys.\ Rev.\ Lett.\ {\bf 99}, 166802 (2007).

\bibitem{Shytov:2007a}
  A.~V.~Shytov, M.~I.~Katsnelson, and L.~S.~Levitov,
  Phys.\ Rev.\ Lett.\ {\bf 99}, 236801 (2007).

\bibitem{Biswas:2007}
  R.~R.~Biswas, S.~Sachdev, and D.~T.~Son,
  Phys.\ Rev.\ B {\bf 76}, 205122 (2007).

\bibitem{Fogler:2007}
  M.~M.~Fogler, D.~S.~Novikov, and B.~I.~Shklovskii,
  Phys.\ Rev.\ B {\bf 76}, 233402 (2007).

\bibitem{Terekhov:2008}
  I.~S.~Terekhov, A.~I.~Milstein, V.~N.~Kotov, and O.~P.~Sushkov,
  Phys.\ Rev.\ Lett.\ {\bf 100}, 076803 (2008).

\bibitem{Kotov:2008a}
  V.~N.~Kotov, B.~Uchoa, and A.~H.~Castro~Neto,
  Phys.\ Rev.\ B {\bf 78}, 035119 (2008).

\bibitem{Kotov:2008b}
  V.~N.~Kotov, V.~M.~Pereira, and B.~Uchoa,
  Phys.\ Rev.\ B {\bf 78}, 075433 (2008).

\bibitem{Pereira:2008}
  V.~M.~Pereira, V.~N.~Kotov, and A.~H.~Castro~Neto,
  Phys.\ Rev.\ B {\bf 78}, 085101 (2008).

\bibitem{Kotov:2012}
  For a comprehensive review, see Sec.~IV in
  V.~N.~Kotov, B.~Uchoa, V.~M.~Pereira, F.~Guinea, and A.~H.~Castro~Neto,
  Rev.\ Mod.\ Phys.\ {\bf 84}, 1067 (2012).

\bibitem{Maksym:2013}
  P.~A.~Maksym and H.~Aoki,
  J.\ Phys.:\ Conf.\ Ser.\ {\bf 456}, 012026 (2013).

\bibitem{Case:1950}
  K.~M.~Case,
  Phys.\ Rev.\ {\bf 80}, 797 (1950).

\bibitem{MathWorld}
  See, for example, {\it Delta Function} in Wolfram MathWorld:
  \verb|http://mathworld.wolfram.com/DeltaFunction.html|.

\bibitem{Nishida:2012}
  Y.~Nishida and D.~Lee,
  Phys.\ Rev.\ A {\bf 86}, 032706 (2012).

\bibitem{Efimov:1970}
  V.~Efimov,
  Phys.\ Lett.\ B {\bf 33}, 563 (1970).

\bibitem{Gorsky:2014}
  A.~Gorsky and F.~Popov,
  Phys.\ Rev.\ D {\bf 89}, 061702(R) (2014).

\bibitem{Bulycheva:2014}
  K.~M.~Bulycheva and A.~S.~Gorsky,
  Usp.\ Fiz.\ Nauk {\bf 184}, 182 (2014)

\bibitem{Novikov:2007}
  D.~S.~Novikov,
  Phys.\ Rev.\ B {\bf 76}, 245435 (2007).

\bibitem{Abramowitz:1965}
  See, for example, {\it Handbook of Mathematical Functions},
  edited by M.~Abramowitz and I.~A.~Stegun (Dover, New York, 1965),
  Chap.~13.

\bibitem{Shytov:2007b}
  A.~V.~Shytov, M.~I.~Katsnelson, and L.~S.~Levitov,
  Phys.\ Rev.\ Lett.\ {\bf 99}, 246802 (2007).

\bibitem{Shytov:2009}
  A.~Shytov, M.~Rudner, N.~Gu, M.~Katsnelson, and L.~Levitov,
  Solid State Commun.\ {\bf 149}, 1087 (2009).

\bibitem{Semenoff:1984}
  G.~W.~Semenoff,
  Phys.\ Rev.\ Lett.\ {\bf 53}, 2449 (1984).

\bibitem{DiVincenzo:1984}
  D.~P.~DiVincenzo and E.~J.~Mele,
  Phys.\ Rev.\ B {\bf 29}, 1685 (1984).

\bibitem{Deshpande:2012}
  A.~Deshpande and B.~J.~LeRoy,
  Physica E {\bf 44}, 743 (2012).

\end{thebibliography}
\end{document}